\documentclass[plb,onecolumn,nofootinbib,showpacs,floatfix,12pt]{revtex4}
\usepackage{graphicx}
\usepackage{amsmath,amssymb,bm,dcolumn}
\usepackage[export]{adjustbox}
\usepackage{graphicx}
\usepackage{float}
\usepackage{amsmath}
\usepackage{amsfonts}
\usepackage{color}
\usepackage{slashed}
\usepackage[colorlinks,hyperindex]{hyperref}
\usepackage{epsfig}		
\usepackage{dcolumn}
\usepackage{bm}
\usepackage{multirow}
\usepackage{longtable}
\usepackage{array}
\usepackage{booktabs}
\usepackage{array} 
\usepackage{varwidth} 

\hypersetup{
colorlinks,
citecolor=blue,
linkcolor=red,
urlcolor=black,
}
\def\be{\begin{equation}}
\def\ee{\end{equation}}
\def\bed{\begin{description}}
\def\eed{\end{description}}
\def\bea{\begin{eqnarray}}
\def\eea{\end{eqnarray}}

\def\ba{\begin{array}}
\def\ea{\end{array}}

\def\u1{$U(1)$}
\def\suu1{$SU(2)\times U(1)$}

\def\0{\mbox{\tiny $0$}}
\def\1{\mbox{\tiny $1$}}
\def\2{\mbox{\tiny $2$}}
\def\3{\mbox{\tiny $3$}}
\def\4{\mbox{\tiny $4$}}
\def\5{\mbox{\tiny $5$}}
\def\6{\mbox{\tiny $6$}}
\def\7{\mbox{\tiny $7$}}
\def\8{\mbox{\tiny $8$}}
\def\9{\mbox{\tiny $9$}}

\def\f14{\mbox{\tiny $\frac{1}{4}$}}

\DeclareMathOperator{\sech}{sech}

\newcolumntype{L}[1]{>{\raggedright\let\newline\\\arraybackslash\hspace{0pt}}m{#1}}
\newcolumntype{C}[1]{>{\centering\let\newline\\\arraybackslash\hspace{0pt}}m{#1}}
\newcolumntype{R}[1]{>{\raggedleft\let\newline\\\arraybackslash\hspace{0pt}}m{#1}}

\begin{document}

\title{Configurational Entropy as a tool to select a physical Thick Brane
Model}
\author{M. Chinaglia$^{1,}$}
\email{chinaglia.mariana@gmail.com}
\author{W. T. Cruz$^{2,}$}
\email{wilamicruz@gmail.com}
\author{R. A. C. Correa$^{1,}$}
\email{fis04132@gmail.com}
\author{W. de Paula$^{1,}$}
\email{wayne@ita.br}
\author{P. H. R. S. Moraes$^{1,}$}
\email{moraes.phrs@gmail.com}
\affiliation{$^1$ITA, Instituto Tecnol\'ogico de Aeron\'autica, 12228-900, S\~ao Jos\'e
dos Campos, SP, Brazil}
\affiliation{$^2$Instituto Federal de Educa\c{c}\~ao Ci\^encia e Tecnologia do Cear\'a, 
\textit{{campus} Juazeiro do Norte, CE, Brazil}}
\date{\today}

\begin{abstract}
We analise braneworld scenarios via a configurational entropy (CE)
formalism. Braneworld scenarios have drawn attention mainly due to the fact
that they can explain the hierarchy problem and unify the fundamental forces
through a symmetry breaking procedure. Those scenarios localize matter in a $%
(3+1)$ hypersurface, the brane, which is inserted in a higher dimensional
space, the bulk. Novel analytical braneworld models, in which the warp
factor depends on a free parameter $n$, were recently released in the
literature. In this article we will provide a way to constrain this
parameter through the relation between information and dynamics of a system
described by the CE. We demonstrate that in some cases the CE is an
important tool in order to provide the most probable physical system among
all the possibilities. \textcolor{black}{In addition, we show that the highest CE is correlated to a tachyonic sector of the configuration, where the solutions for the corresponding model are dynamically unstable.}

\end{abstract}

\date{\today}
\maketitle



\section{Introduction}

Due to the inability revealed by General Relativity (GR) in accurately
describe some cosmological and astrophysical issues \cite%
{padmanabhan/2003,bull/2016,sahni/2004,oman/2015,sahni/2002}, new theories {%
of gravity} have been proposed. Braneworld {models are a possibility} have
been in vogue in the last decades because when extra dimensions are
considered, they lead to modified field equations that can handle some of
those issues in a healthy way. For example, brane scenarios allow for
neutron star masses around 2 solar masses, in agreement with observations.
Those limits cannot be attained by GR, unless an exotic equation of state is
introduced. Besides, this scenarios could be a properly way to solve the
hierarchy problem \cite{randall/1999,arkani-hamed/1998}.

Novel analytical models derived in \cite{branemariana} describe the universe
through the brane paradigm and have the prerogative of depending on a free
parameter $n$. Although varying $n$ leads to different characteristics in
comparison to the literature until then, this parameter has never been
constrained by any physical {approach}.

\textcolor{black}{However, a few years ago, Gleiser and Stamatopoulos, motivated by the Shannon's information theory \cite{shannon}, have introduced in the literature the so-called configurational entropy (CE) concept \cite{gleiser}. This new physical quantity, which is a measure in the functional space, is able to relate the dynamical and informational content of physical models with localized energy configurations, allowing the establishment of optimized analytical solutions in nonlinear field theories. CE also can be used to resolve situations where the configurations present degeneracy \cite{rafael-gleiser}, providing us a complementary perspective to understand situations where arguments based in the energy analysis are inconclusive.}

\textcolor{black}{Interesting applications of the CE approach can be found in several areas, such as the spontaneous symmetry breaking context \cite{gleiser-niki}, Q-balls structures \cite{gleiser-sow}, the study of self-gravitating astrophysical objects \cite{gleiser-jiang},  the critical behavior of continuous phase transitions in the context of (2+1)-dimensional Ginzburg-Landau model \cite{gleiserdamian}, the framework of the Color-Glass Condensate (b-CGC) dipole model \cite{gayane},  anti-de Sitter black holes \cite{nelson}, dynamical AdS/QCD holographic models \cite{alex,wayne1,wayne2,wayne3} and charged AdS black holes \cite{lee}.}

\textcolor{black}{Another important scenario, where CE plays a fundamental role, is in Lorentz and CPT breaking systems \cite{rafael-LV}. In this context, it is possible to determine the bounds for the Lorentz violation parameters, which opens a new window to experimental investigations in theories with Lorentz symmetry breaking.}

\textcolor{black}{CE has also been shown to be very useful in restricting braneworld parameters. We can think of our universe as a (3+1) dimensional space-time, regarded as a brane, embedded in a higher-dimensional bulk, where the extra dimension can be large or compact. Within this scope, it was shown by Correa and Rocha \cite{correa-rocha} that information entropic measure is an accurate way for providing the most suitable values for the bulk AdS curvature. Following that work, we can find different lines of investigations in braneworld models, in which the CE allows one to obtain important results. Such studies arise in $f(R)$ \cite{correa-dutra-rocha} and $f(R,T)$ \cite{correa} theories, in Weyl \cite{correa-davi} and Gauss-Bonnet brane models \cite{correa-wayne}, in string scenarios \cite{correa-carlos} and in thick branes \cite{correa-wilami}.}

\textcolor{black}{Thus, in the present work} we propose to use the CE \textcolor{black}{concept presented by Gleiser and  Stamatopoulos} \textcolor{black}{in order to investigate} some of the novel analytical braneworld models presented in \cite{branemariana}. \textcolor{black}{Our goal is to} provide a physical constrain on the parameter $n$, \textcolor{black}{which is responsible for generating different classes of configurations}. This analysis will allow us to find the most-likely physical scenario among all possible values that $n$ can assume. \textcolor{black}{Therefore, we will use the informational-entropic measure in order to find which is the most favourable configuration of a given family with same genealogy.  It is important to remark that} a similar analysis was made in \cite{correa} in favor of finding a bound in $f(R,T)$ theories.

The paper is organized as follows: in Section II we describe the braneworld
model {in question}, in Section III we review the mathematical approach for
the CE calculation, in Section IV we present the CE calculation for the
brane scenarios, in Section V we discuss the results and, finally, in
Section VI we present the conclusions.

\section{Brane Models}

\label{brane}

\textcolor{black}{In this section we will look at a braneworld scenario in $(4+1)$ dimensions. We will work with a theory which consists of a scalar field and a manifold with a 4D subspace controlled by a warp factor, where the equations of motion are derived for both. In this case, we show that it is possible to use an ansatz for the solution to the warp factor by invoking cyclic deformation chains. As a consequence, using these analytical solutions, we can write the energy density of these models in the bulk.}

\textcolor{black}{We would like to highlight that} braneworld scenarios were studied through several analytical procedure 
strategies \cite{RSII,Brana2}. Concerning this paradigm, the observable
universe is a manifold with (3+1) dimensions supported by a warp factor
responsible for matter localization inside the brane.

The braneworld models studied in this paper were derived in \cite%
{branemariana} where they were associated with generic solutions of an
effective action driven by a real scalar field, $\phi$, coupled to gravity as in the model of Reference \cite{RSII}. \textcolor{black}{In the case of Ref.\cite{branemariana}, the corresponding action is written as}
\begin{equation}
S=\int{d^4x dy \sqrt{\mid g \mid}\bigg[ -\frac{1}{4}R+\frac{1}{2}%
\partial_{a}\phi\partial^{a}\phi-V(\phi) \bigg]},  \label{acaod}
\end{equation}
where $\mid g\mid=det(g_{ab})$, $R$ is the scalar curvature in 5D, $a$
varies from $0$ to $4$ {and $V(\phi)$ is the scalar field potential.}

\textcolor{black}{In \cite{branemariana},} it was chosen a 5-dimensional braneworld scenario, whose interval is
expressed by
\begin{equation}
ds^2=e^{2A}\eta_{\mu\nu}dx^{\mu}dx^{\nu}-dy^2,  \label{intervaloinvariante}
\end{equation}
{with $e^{2A}$ being the warp factor, $\eta_{\mu\nu}$ is the Minkowski
metric, $\mu,\nu$ run from $0$ to $3$ and $y$ is the extra coordinate.}

Assuming a dependency of $\phi$ exclusively on the extra dimensional
coordinate $y$, \textcolor{black}{i.e., $\phi=\phi(y)$,} the equation of motion for the scalar field arising from the
action (\ref{acaod}) is given by 
\begin{equation}
\frac{d^2\phi}{dy^2} + 4 \frac{d A}{dy} \frac{d \phi}{dy} - \frac{d}{d\phi}%
V(\phi) = 0.  \label{eom}
\end{equation}

\textcolor{black}{On the other hand,} by varying action (\ref{acaod}) with respect to the metric, one obtains 

\begin{equation}
\frac{3}{2}\frac{d^2 A}{dy^2} = - \left(\frac{d\phi}{dy}\right)^2.
\label{eqaphi}
\end{equation}

\textcolor{black}{We would like to remark that} the warp factor, $e^{2A}$, was constrained from the ansatz used in {\cite%
{branemariana}}. Considering that energy density localized structures demand
for a lump-like warp factor behavior, so that matter integrability and
localization are assured on the brane, there can be assumed a
straightforward identification such as $e^{2A}=\psi$, where $\psi$ is the
bell-shaped analytical expression known \emph{a priori}. This assumption
leads to {\cite{branemariana}}

\begin{equation}
A=\frac{1}{2}\ln{\psi}.  \label{A2}
\end{equation}

A plethora of analytical lump solutions was obtained in \cite{alexroldao},
generated via cyclic deformation chains (CDC). They were triggered by the
solution of the $\lambda\chi^4$ {potential} (in addition of other three
deformed models). Chinaglia, Bernardini and Rocha inserted some
of the deformed lump solutions as the ansatz {for} $\psi$ in Eq.~(\ref{A2})
in order to build up novel analytical braneworld scenarios. A complete
review of generic deformations can be found, for example, in \cite{Bas01},
while the CDC procedure was proposed in \cite{alexroldao}. In \cite{Bas01}
the deformation generates a new system in a way such that one increases or
decreases the amplitude and width of the defect, without changing the
corresponding topological behavior. In \cite{alexroldao} the authors derived
deformed defects that, in spite of having different amplitudes and widths,
can interchange from topological (kink) to non-topological (lump) profiles.
Such a behavior can also be seen in \cite{alexmariana}. The deformation
procedure as well as the lump solutions are quite extensive and do not
concerns the scope of this work, however we dedicate a consistent review of
it in Appendix A.

Since the lump solutions derived in \cite{alexroldao} have a dependency on
the parameter $n$, the brane warp factor also presents such a dependency.
Our interest is to make use of the CE \cite{gleiser} as a tool to set the
free parameter $n$. \textcolor{black}{Here, we will use the CE approach with the purpose of establishing the physically acceptable configurations, where a stable brane can be constructed within a positive energy density spectrum. In this case, we are seeking for a correlation between $n$ and minimum values of CE. It is important to remark that finding parameters responsible for the stability of the brane is very important to analyze the issue of localization of fermions on these branes, thus making the CE a necessary tool to constraint such models.} For this purpose, it is essential to know the brane energy density spectrum. This was also calculated in \cite{branemariana}
through the $00$-component of the energy-momentum tensor inside the brane
derived in \cite{A,B}: 
\begin{equation}
T_{00}= \bigg[ \bigg( \frac{d\phi}{dy} \bigg)^2-3 \bigg( \frac{dA}{dy} %
\bigg)^2 \bigg]e^{2A(y)}.  \label{tensor}
\end{equation}

It is also important to stress that from Eq. (\ref{eqaphi}) one can infer
that only some values of $n$ will generate real physical solutions, $\phi$,
as will be discussed in details in Section V.

We now present the warp factor, $e^{2A(y)}$, and the $00$ component of the
energy-momentum tensor, $T_{00}(y)$, for the 4 braneworld models of Ref.\cite%
{branemariana}. Those will be analyzed via CE in Section IV. Each subsequent
model will be referred by the numbers, 1, 2, 3 and 4 following the sequence
they are presented. The behavior {of their warp factor and matter-energy
density} are depicted in Figs.~1, 2, 3 and 4.

The first model considered was driven by the \emph{ansatz}\footnote{%
In order to understand the \textit{ansatz} derivation for all the models,
see Appendix A.}: 
\begin{equation}
\psi_1=\frac{-\ln[{\cosh[n \tanh(y)] \sech(n)}]}{n},
\end{equation}
which inserted in Eq. (\ref{A2}) provide the function $A(y)$ and this
enables the energy-momentum calculation: 
\begin{eqnarray}
A(y)_1 &=& \frac{1}{2} \ln \bigg[\frac{-\ln[{\cosh[n \tanh(y)] \sech(n)}]}{n} %
\bigg], \\
T_{00}(y)_1 &=& \frac{3}{4} \bigg[ n \sech(y)^4 \sech[n\tanh(y)]^2 -2 \sech%
(y)^2\tanh(y)\tanh[n \tanh(y)] \bigg] \, ,
\end{eqnarray}
that are represented in Fig.~1. 

\begin{figure}[h!]
\centering
\textbf{Model 1}
\par
\medskip \includegraphics[scale=0.57]{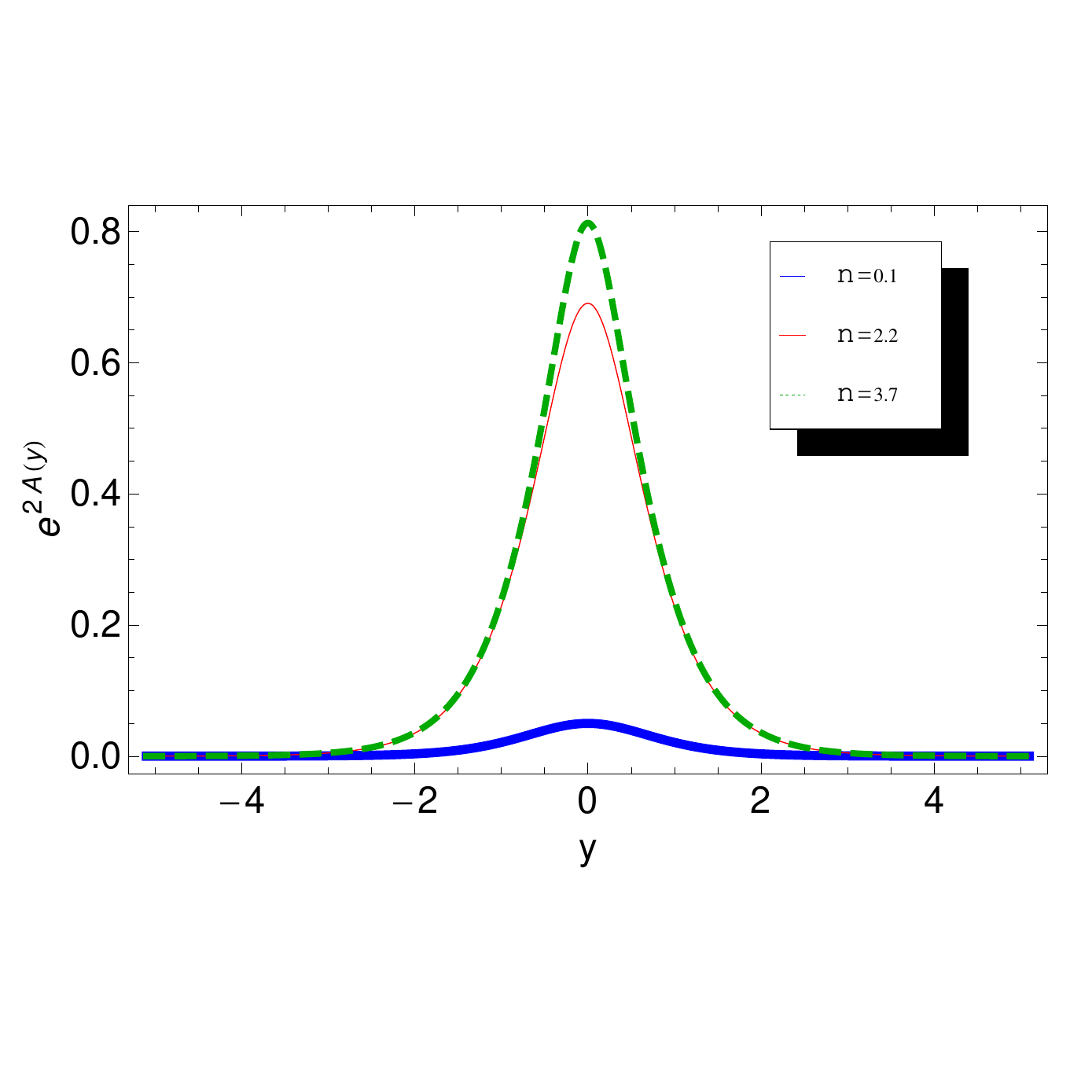}
\includegraphics[scale=0.57]{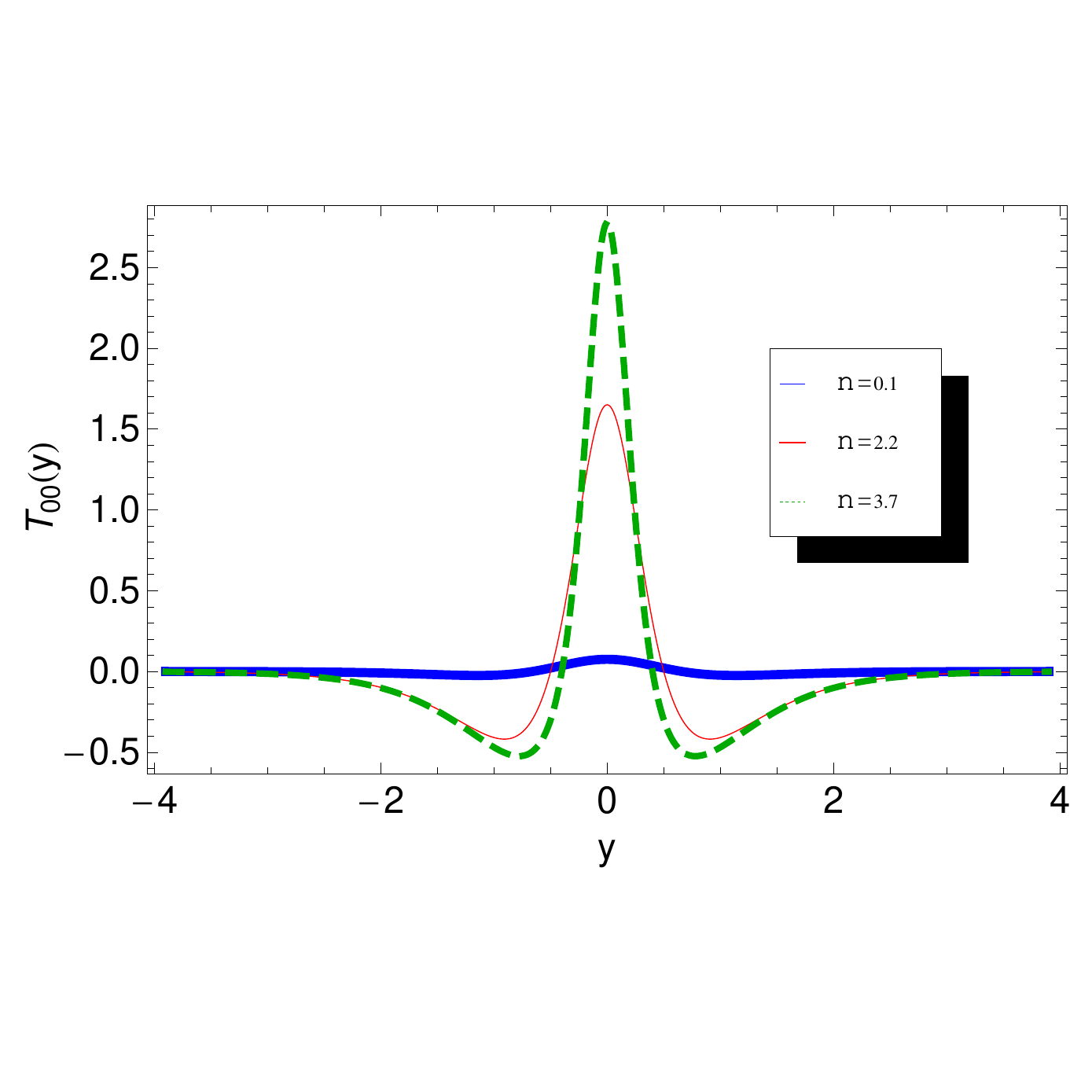}
\caption{\textcolor{black}{Sample profiles of the warp factor (left) and energy-momentum tensor (right) for Model 1.
The thick lines correspond to the case with $n=0.1$, the thin continuous lines depict the profiles for $n=2.2$ while the dashed lines represent the case with $n=3.7$.}}
\label{modelo1}
\end{figure}

Fig.~2 presents the second model obtained from the \emph{ansatz}: 
\begin{equation}
\psi_2=\frac{\cos[n \tanh(y)] - \cos(n)}{n},
\end{equation}
that generates the scenario with: 
\begin{eqnarray}
A(y)_2 &=& \frac{1}{2} \ln[{\frac{-\cos(n) + \cos[n \tanh(y)]}{n}}], \\
T_{00}(y)_2 &=& \frac{3}{4} \bigg[ n \cos[n \tanh(y)] \sech(y)^4 - 2 \sech%
(y)^2 \sin[n \tanh(y)] \tanh(y) \bigg].
\end{eqnarray}

\begin{figure}[h!]
\centering
\textbf{Model 2}
\par
\medskip \includegraphics[scale=0.57]{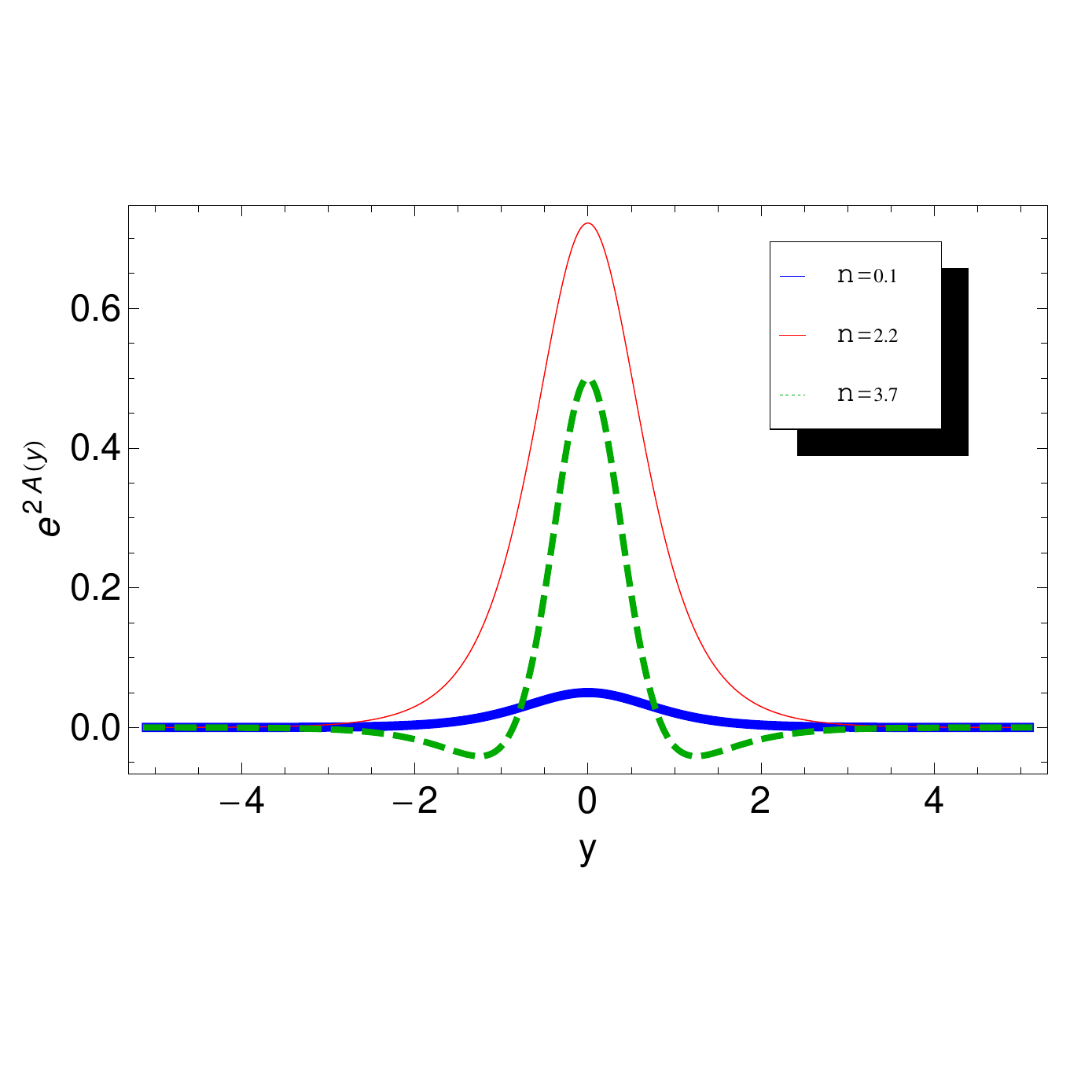}
\includegraphics[scale=0.57]{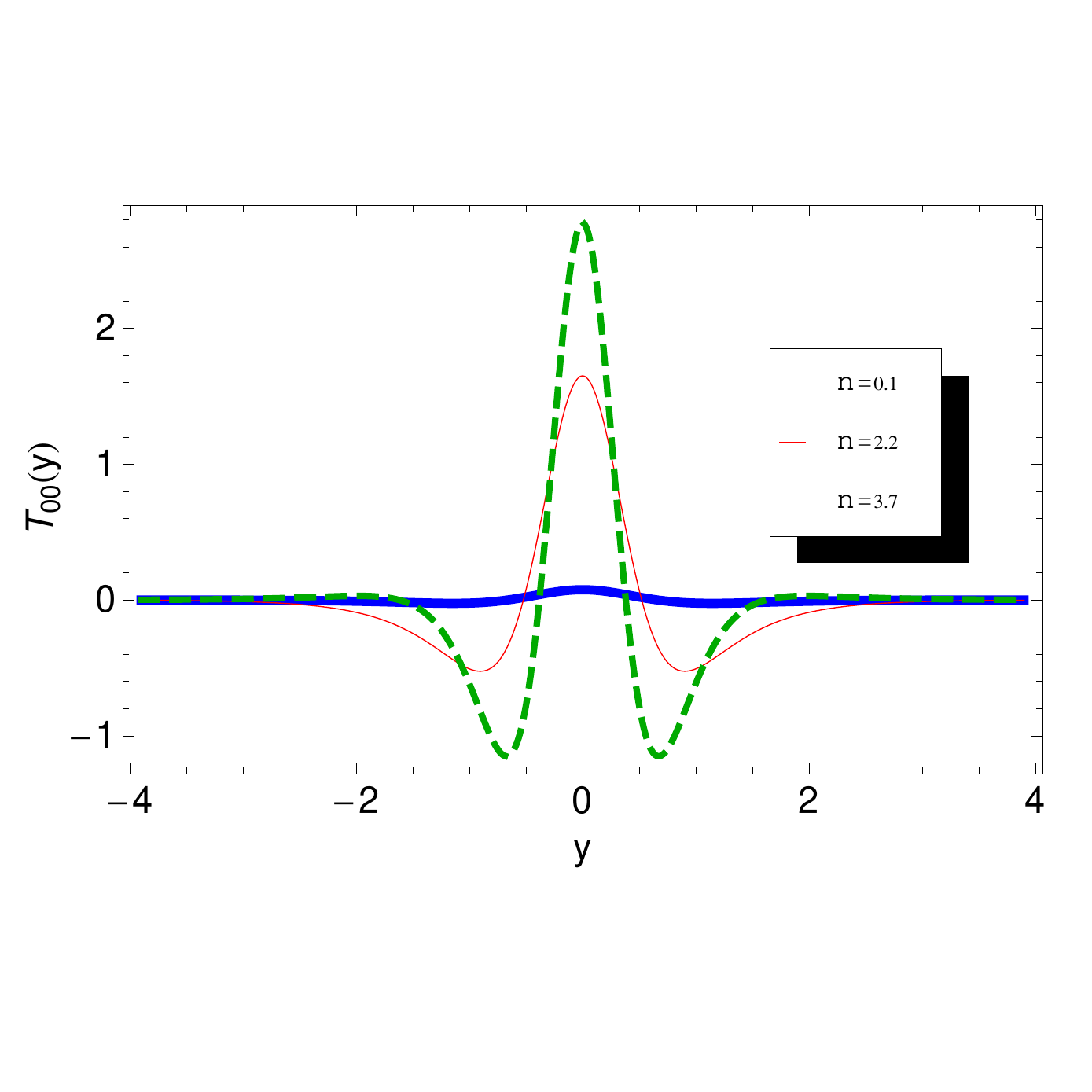}
\caption{\textcolor{black}{Sample profiles of the warp factor (left) and energy-momentum tensor (right) for Model 2.
The thick lines correspond to the case with $n=0.1$, the thin continuous lines depict the profiles for $n=2.2$ while the dashed lines represent the case with $n=3.7$.}}
\label{modelo2}
\end{figure}

Model 3 has been derived from the ansatz: 
\begin{equation}
\psi_3 =\frac{(\sech[n \tanh(y)] - \sech (n))}{n}
\end{equation}
which leads to: 
\begin{equation}
A(y)_3 = \frac{1}{2} \ln{\left[\frac{\sech [n \tanh(y)]-\sech (n)}{n}\right]}%
,
\end{equation}
\begin{equation}
\begin{split}
T_{00}(y)_3 = -\frac{3}{8}\sech (y)^3\sech [n \tanh(y)]^3 \times \\
[n(-3+\cosh(2n \tanh(y)))\sech (y)+2\sinh(y)\sinh(2n\tanh(y))],
\end{split}%
\end{equation}
whose behavior is depicted in Fig.~3.

\begin{figure}[h!]
\centering
\textbf{Model 3}
\par
\medskip \includegraphics[scale=0.57]{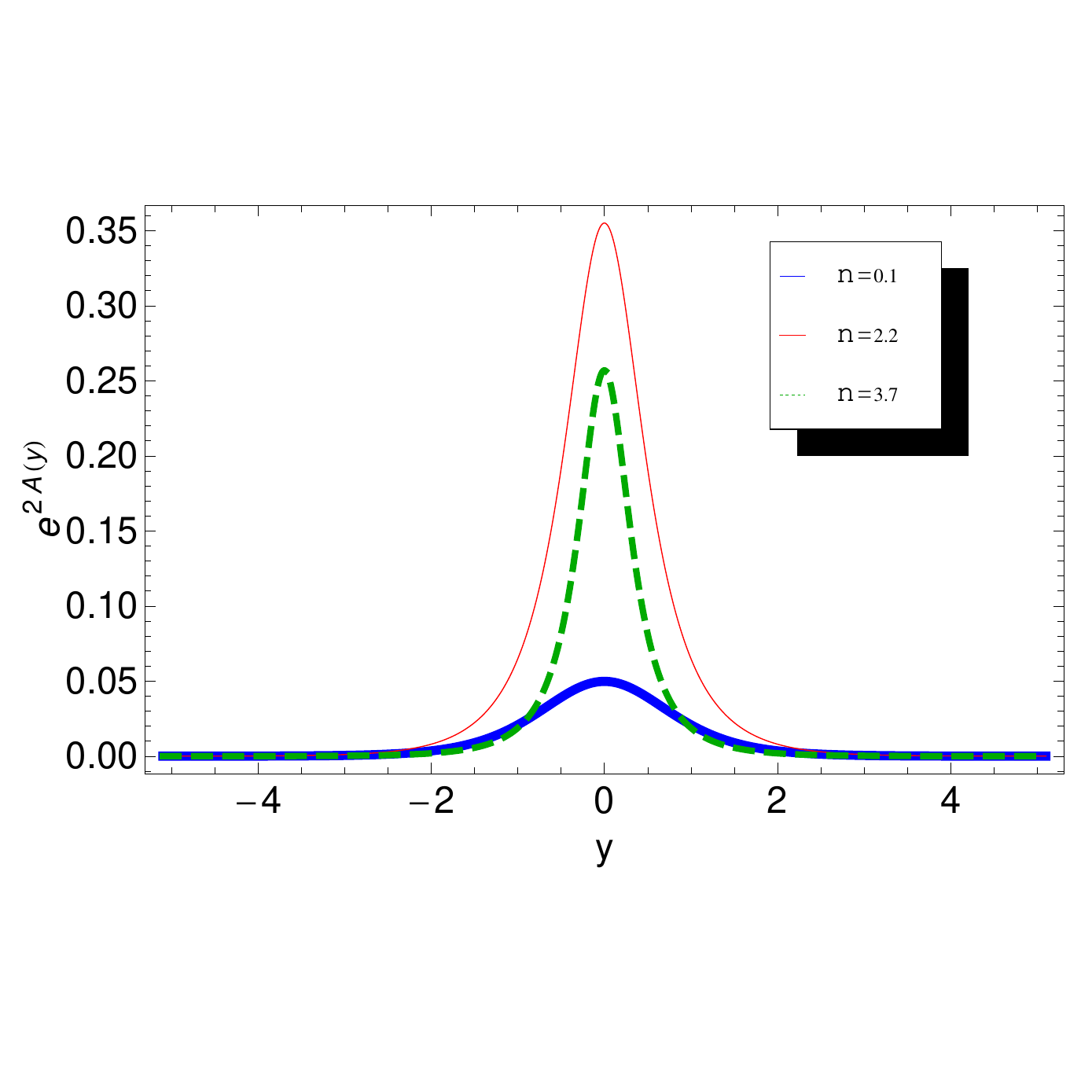}
\includegraphics[scale=0.57]{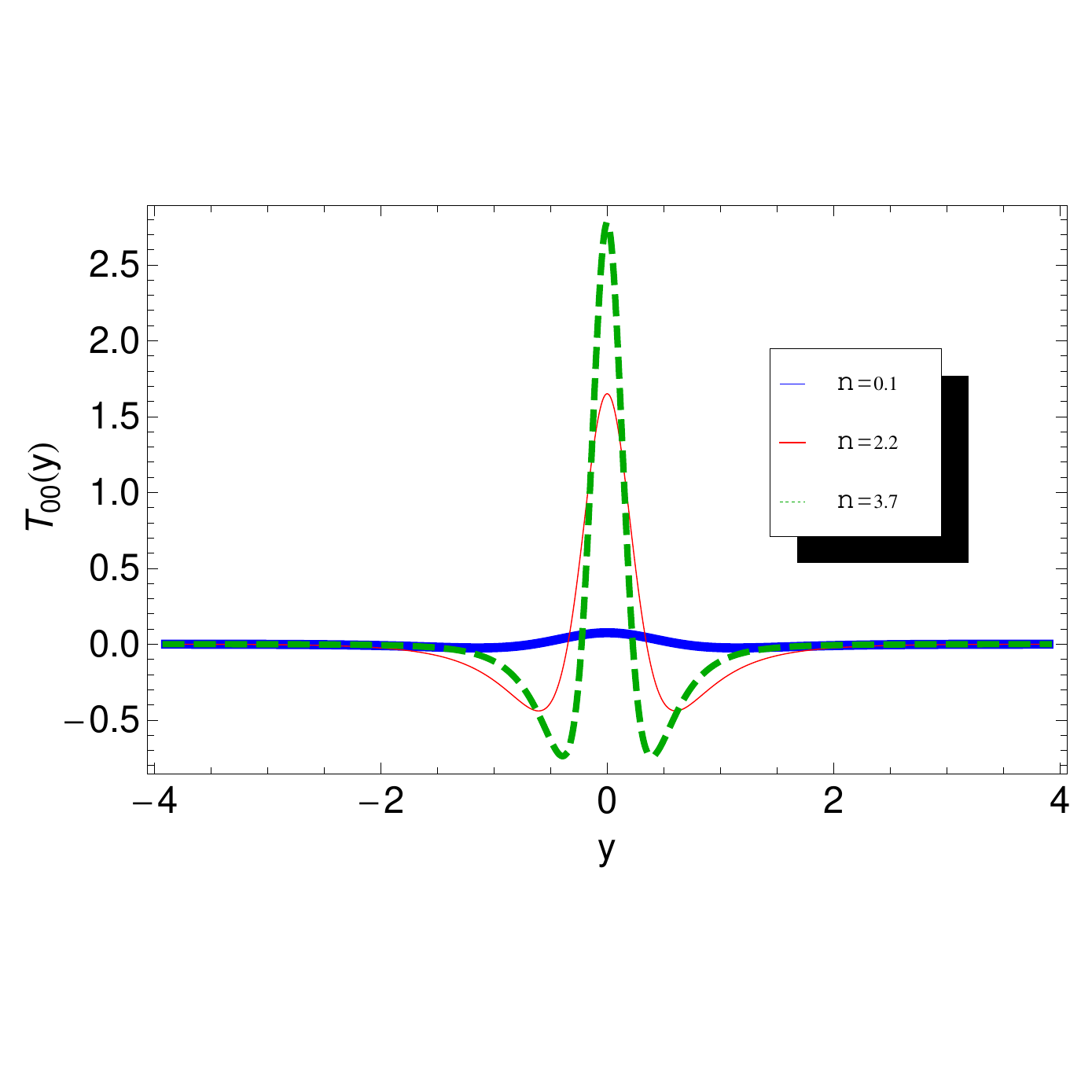}
\caption{\textcolor{black}{Sample profiles of the warp factor (left) and energy-momentum tensor (right) for Model 3.
The thick lines correspond to the case with $n=0.1$, the thin continuous lines depict the profiles for $n=2.2$ while the dashed lines represent the case with $n=3.7$.}}
\label{modelo3}
\end{figure}

Finally, Model 4 was built through the ansatz: 
\begin{equation}
\psi_4 = \frac{(2 n \sech(y) + \sin[2 n \sech(y)])}{4 n},
\end{equation}
that engenders: 
\begin{equation}
A(y)_4 = \frac{1}{2} \ln \bigg[\frac{2 n\, \sech(y) + \sin[2 n \,\sech(y)]}{%
4 n} \bigg] \\
\end{equation}

\begin{equation}
\begin{split}
T_{00}(y)_4 =\frac{3}{4}\sech(y)[\cos(n \sech(y))^2\sech(y)^2+ \\
\tanh(y)^2(-\cos[n\sech(y)]^2+n\sech(y)\sin[2n\sech(y)])].
\end{split}%
\end{equation}
as can be seen in Fig.~4.

\begin{figure}[h!]
\centering
\textbf{Model 4}
\par
\medskip \includegraphics[scale=0.57]{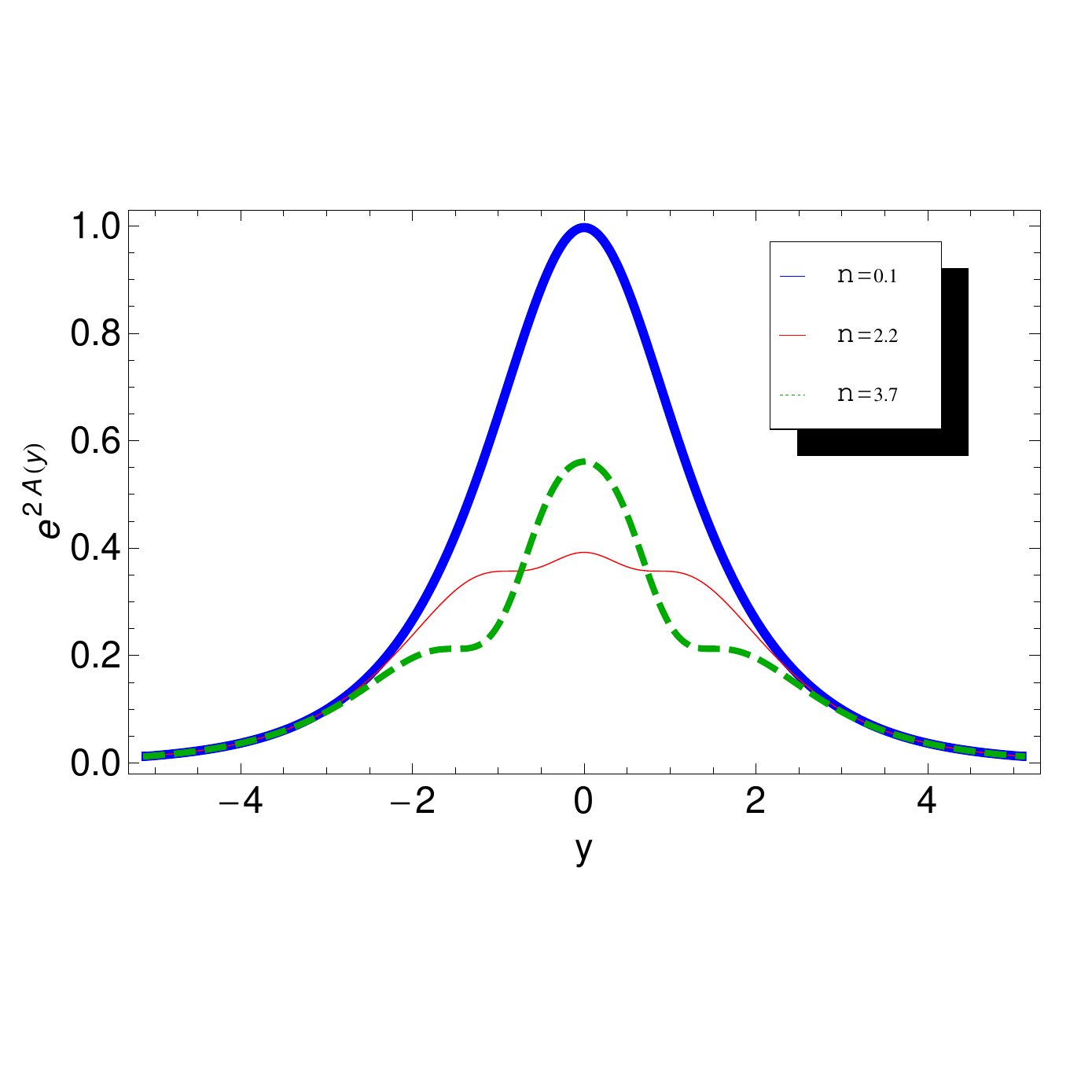}
\includegraphics[scale=0.57]{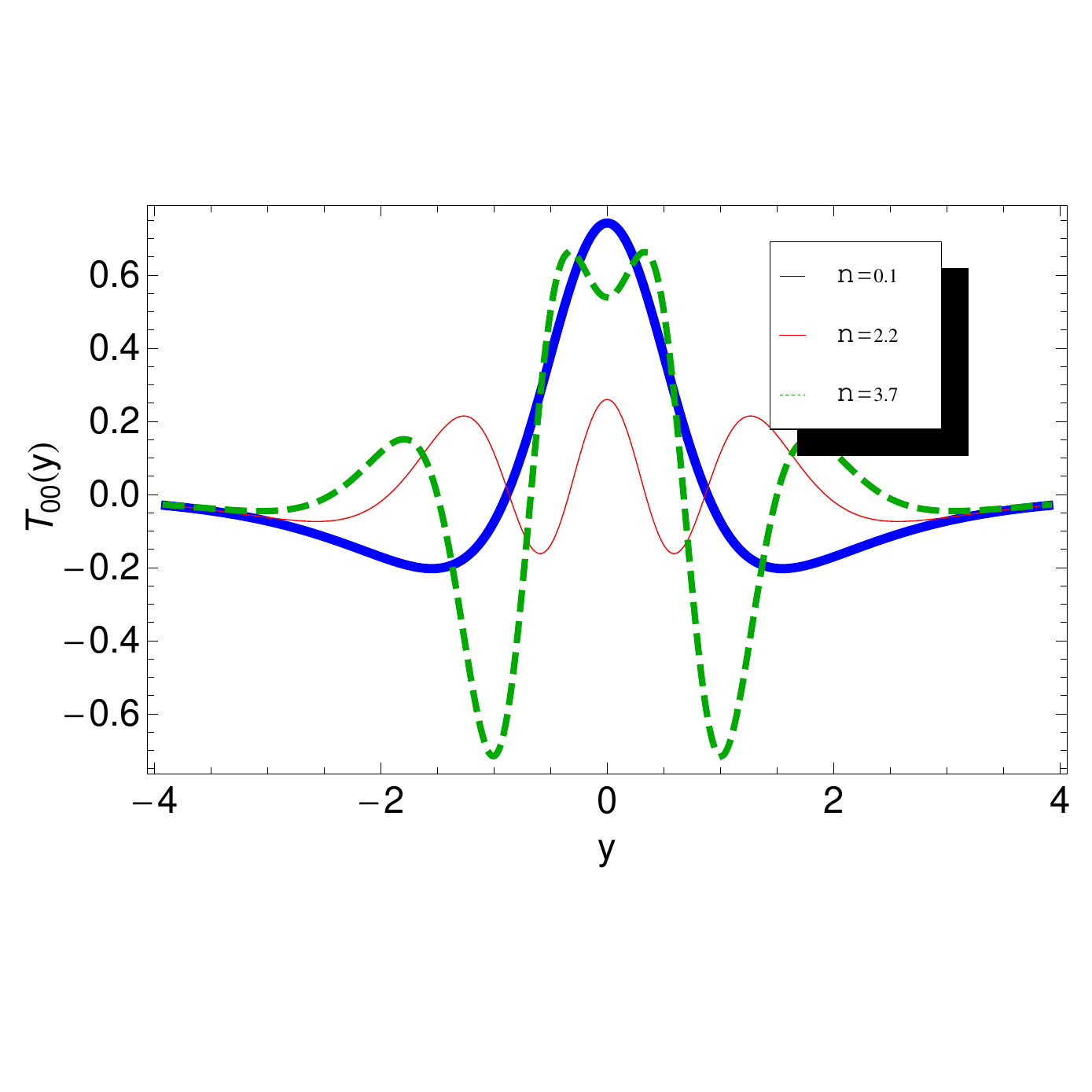}
\caption{\textcolor{black}{Sample profiles of the warp factor (left) and energy-momentum tensor (right) for Model 4.
The thick lines correspond to the case with $n=0.1$, the thin continuous lines depict the profiles for $n=2.2$ while the dashed lines represent the case with $n=3.7$.}}
\label{modelo4}
\end{figure}

\textcolor{black}{From Figs. 1, 2, 3 and 4, we can see that the warp factor and the energy density are mostly concentrated at the origin of the fifth dimension. In the next Section, we will apply the CE procedure in order to find the parameters $n$ which minimize the information entropic. Then, we will show the correlation between dynamical and information content of these models.}

\section{Configurational Entropy}

\textcolor{black}{The CE concept was defined a few years ago by Gleiser and Stamatopoulos \cite{gleiser}. Such a new physical quantity is a Shannon-like entropic measure, which can be used to study nonlinear scalar field models where there are some dubious information regarding the field configurations and their energies. In such cases, the CE is a fundamental physical quantity which should provide important bounds and constraints about the solutions. In addition, when there are degenerate energies, the information-entropic measure is a powerful approach to break degeneracy \cite{rafael-gleiser} and point out the most favourable configuration within the numerous possibilities offered by the system.}

\textcolor{black}{In this Section, for convenience and in order to put our work in a more pedagogical form, let us show a review of the basic concepts on CE in brane world scenarios \cite{correa-rocha} and its consequences.}
 
\textcolor{black}{Let us begin by considering the brane localized energy density pattern, which is given by the energy-momentum tensor $T_{00}(y)$. Thus, it is possible to consider a set of square-integrable bounded functions $T_{00}(y)$ $\in$ $L^2(\Re)$ and their Fourier transforms $P(k)$, where using the Plancherel's theorem it follows that}

\begin{equation}
\int_{-\infty}^{+\infty} |T_{00}(y)|^2dy=\int_{-\infty}^{+\infty}|P(k)|^2dk.
\end{equation}
\textcolor{black}{Now, following the Ref. \cite{gleiser}, the so-called modal fraction is constructed from the normalized power
spectrum of the field and it is defined by the following expression}
\begin{equation}  \label{modal}
f(k)=\frac{|P(k)|^2}{\int|P(k)|^2dk }.
\end{equation}

\textcolor{black}{We highlight that the modal fraction $f(k)$ measures the relative weight of each mode $k$. On the other hand,} if the functions are periodic, they can be written as Fourier series. In this case, each squared term represents the probability of occurrence of the 
$k$ mode, such that $f(k)\rightarrow f_n= |P_n|^2/\sum |P_n|^2$. \textcolor{black}{Moreover, it is important to remark that the modal fraction plays the role of the probability distribution since it gives the relative contribution of a given mode $k$ for the CE.} .

We define the CE as \cite{gleiser}
\begin{equation}  \label{EC}
S_C[f]=-\sum f_n \ln{(f_n)}.
\end{equation}
Note that Eq. (\ref{EC}) is analogue to the Shannon informational entropy,
which represents a limit for the best way to compactify information without
losses \cite{shannon}. Besides, note that both Eq. (\ref{EC}) and
Shannon's entropy are also analogue to the usual thermodynamical entropy
described by Boltzmann equation, $S = -R f \ln (f)$, considering the
Stirling approximation: $\ln f! \approx f\ln f$. Here, $f$ is the frequency of some
event divided by the number of total events. As it is defined, CE provides
the informational content of the possible configurations compatible with the
constrains of the physical system. If there are constrains on the system, a
shorter amount of information is reacquired. If there is no constrain, the
system is completely aleatory and all the information will be relevant.

In order to find such constrains, Shannon defined a redundancy factor in
terms of a maximum entropy value. This maximum condition is attained when
the uncertainty of the event is also maximal (completely aleatory event), so
that the informational content intrinsic to each event is higher. In order
to acquire a maximum value of uncertainty, one must have the most equal
distribution as possible for the frequency associated to each event. That
is, all the events must have the same probability, so that the entropy will
be maximized.

\textcolor{black}{Using a connection with information theory, one way to understand the CE concept is to consider that the message sent by a given system will be its configuration, which must be represented by a spatially-bounded or periodic function. On the other hand, the alphabet corresponds to the momentum modes which compose the configuration with specific weights, that is to say probabilities, where such modes are obtained applying a Fourier transform from the system solutions. Therefore, the configuration (message) assumes a specific informational signature in momentum space establishing a quantifiable complexity. Thus, it can be concluded that the CE measures the relative spatial complexity of the function describing the physical system in terms of its momentum-mode decomposition.}

Note that, even if the function has any periodicity, it is still possible to
define the \textcolor{black}{continuous} CE
\begin{equation}  \label{CE}
S_c[f]=-\int \tilde{f}(k) \ln [\tilde{f}(k)]dk
\end{equation}

\textcolor{black}{In the next Section we will use the CE approach} in order to find
the constrains for the brane scenario presented in Section II. \textcolor{black}{As we will see,} when all the $k$ modes carry the same weight (i.e., the same probability), the CE is maximum. By doing this,
we will be able to compare both the real and the maximum entropy to find the
constrains and, consequently, the redundancy. We will then constrain the
free parameter $n$. Consequently, this work will provide a physical
observable, which is capable of selecting the most probable
physical model.

\section{Configurational entropy for the brane scenarios}

\begin{figure}[h!]
\centering
\includegraphics[scale=0.33]{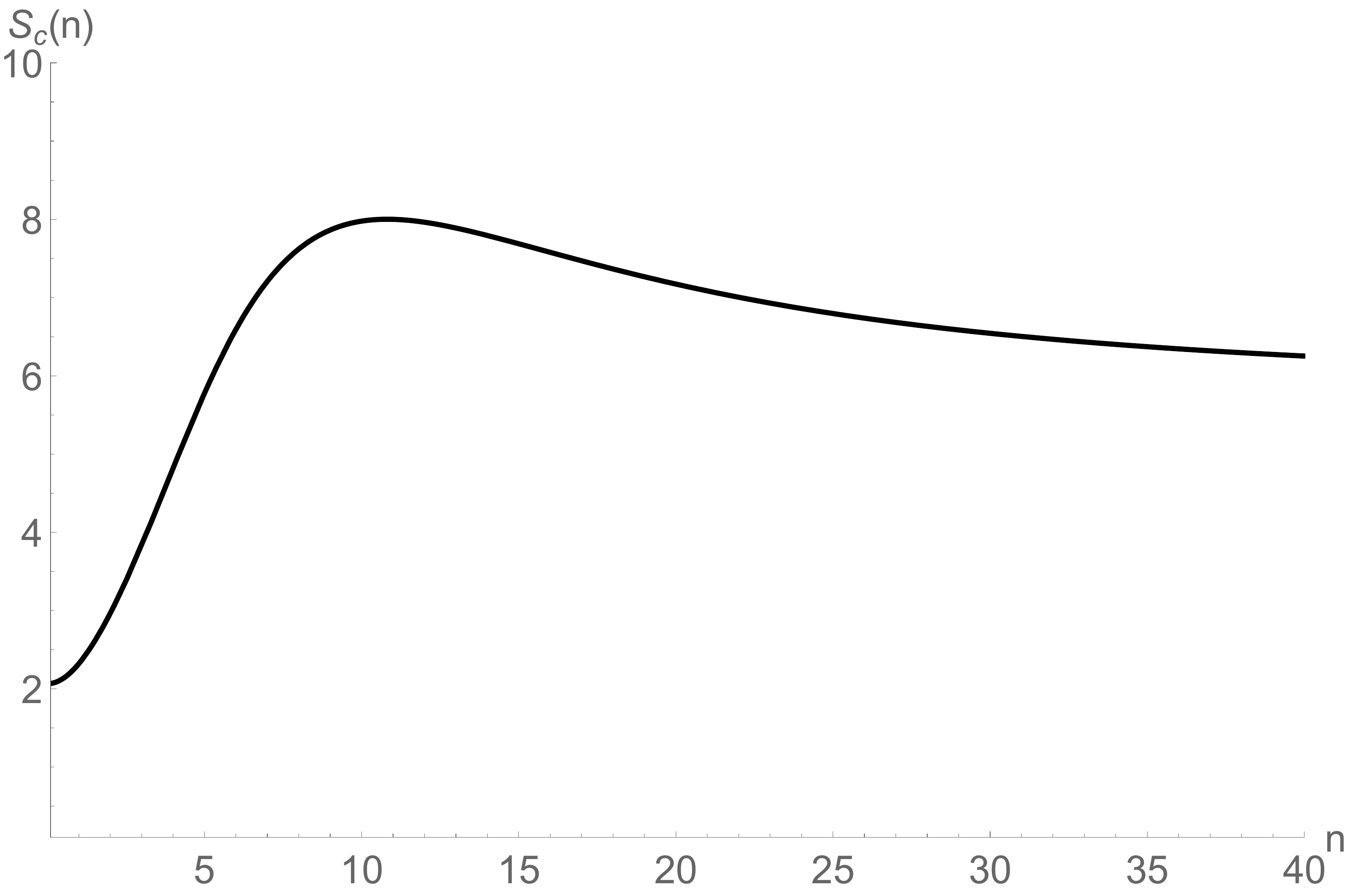}
\caption{CE for the Model 1 with $0.1\leqslant n\leqslant40$. The minimum CE
is 2.0665 at $n=0.1$ and the maximum CE is 8.0008 at $n=10.8161$.}
\label{model1}
\end{figure}

In this section we evaluate the CE resulting from each brane model presented
in Sec.\ref{brane}. In order to achieve the modal fraction (\ref{modal}), we
need to know the energy density along the extra dimension. Due to the
complexity of the solutions for $T_{00}(y)$, we numerically calculate the
Fourier transform and from there we obtain the modal fraction. \textcolor{black}{In this case, the Fourier transform assumes a high oscillatory integrand when increasing $k$. The numerical calculations were made in Mathematica software and we have used  an extrapolating oscillatory integration strategy, since we have oscillating integrals in infinite one-dimensional regions. As the rule of integration we have adopted the Levin-type rule. The CE integral was evaluated repeatedly in terms of the parameter $n$ varying with $0.1$ step.} To finally
obtain the CE from Eq. (\ref{CE}) we define the normalized modal fraction as 
$\tilde{f}(k)=f(k)/f_{max}(k)$, \textcolor{black}{which guarantees the positivity of the CE}. The integration region to the CE in
Eq. (\ref{CE}) is chosen so that at its limits we obtain $\tilde{f}(k)
\ln [\tilde{f}(k)]\rightarrow0$. Our results are presented as the plots
of the CE in terms of the parameter $n$ in each case. For convenience we
choose the warp factor so that $e^{2A(0)}=1$. Then, we adopt $\tilde{A}%
(y)=A(y)-A(0)$ and find new expressions for $T_{00}(y)$. For Model 1, we obtain the following energy density from the normalized warp factor $\tilde{A}(y)$ 
\begin{equation}
\rho(y)_1 =- \frac{3 n \sech(y)^3\big[n \sech(y) \sech[n \tanh(y)]^2-2
\sinh(y) \tanh[n \tanh(y)]\big]}{4 [\ln\sech(n)]}.
\end{equation}
The corresponding CE is presented in Fig. \ref{model1}. The lowest value
obtained for the CE was 2.0665. When $n$ acquires large values, the CE tends
to 5.8331. 

\begin{figure}[h!]
\centering
\includegraphics[scale=0.29]{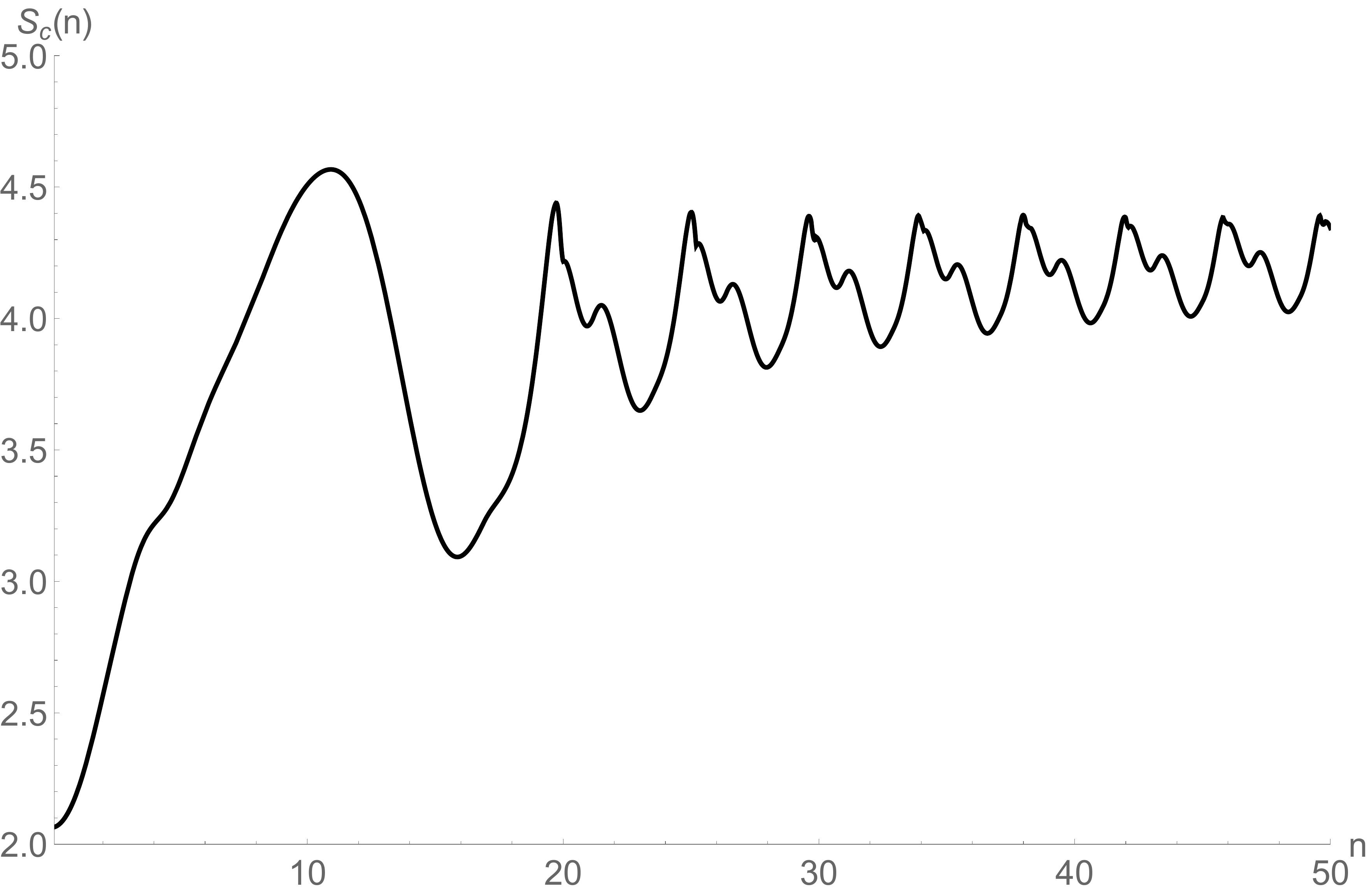}
\caption{CE for the Model 2 with $0.1\leqslant n\leqslant50$. The minimum CE
is 2.0651 at $n=0.1$ and the maximum CE is 4.5671 at $n=10.9191$.}
\label{model2}
\end{figure}

For Model 2 we find the following energy density with the normalized warp
factor: 
\begin{equation}
\rho(y)_2 =-\frac{3n \sech(y)^3 \big[n\cos[n \tanh(y)] \sech(y) - 2 \sin[n
\tanh(y)] \sinh(y)\big]}{ 4 [-1 + \cos(n)]}.
\end{equation}
The corresponding CE is plotted in Fig. \ref{model2}.  

\begin{figure}[h!]
\centering
\includegraphics[scale=0.29]{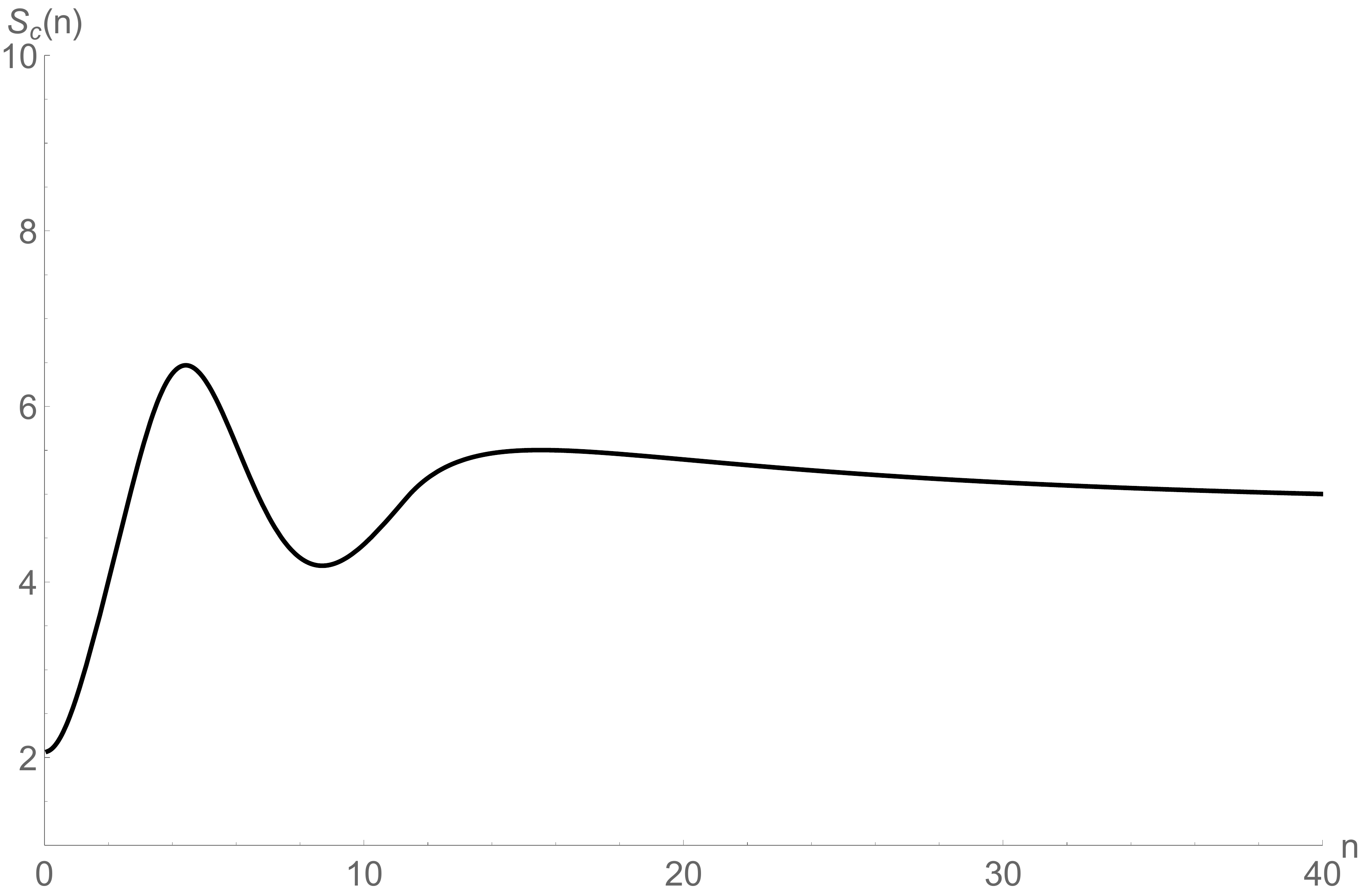}
\caption{CE for the Model 3 with $0.1\leqslant n\leqslant40$. The minimum CE
is 2.0706 at $n=0.1$ and the maximum CE is 6.4687 at $n=4.4245$.}
\label{model3}
\end{figure}

The Model 3 have the following energy density: 
\begin{equation}
\begin{split}
\rho(y)_3=\big[n (-3 + \cosh[2 n \tanh(y)]) \sech(y) + 2 \sinh(y) \sinh[2 n
\tanh(y)]\big]\times \\
\frac{3 n \sech[y]^3 \sech[ n \tanh (y)]^3 }{8[1-\sech(n)]}
\end{split}%
,
\end{equation}
and its corresponding CE is plotted in Fig. (\ref{model3}).

\begin{figure}[h!]
\centering
\includegraphics[scale=0.29]{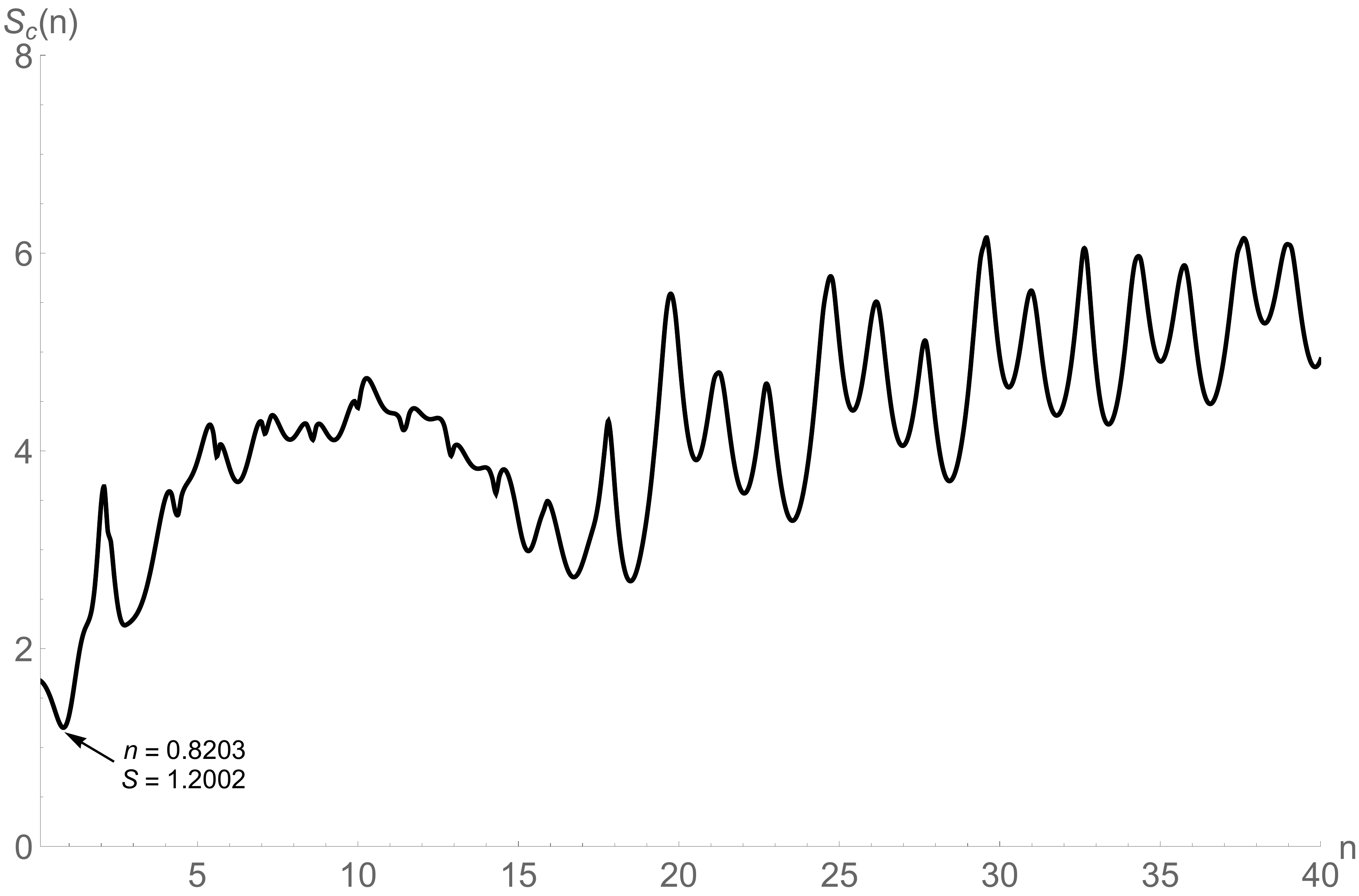}
\caption{CE for the Model 4 with $0.1\leqslant n\leqslant40$. The minimum CE
is 1.2002 at $n=0.8203$.}
\label{model4}
\end{figure}

At last, the energy density coming from the normalized warp factor for Model
4 is 
\begin{equation}
\begin{split}
\rho(y)_4=\frac{3 n \sech( y)^3 \big[-\cos[n \sech(y)]^2 (-1 + \sinh(y)^2) +
n \sin[2 n \sech(y)] \sinh(y) \tanh(y)\big]}{[2 n + \sin(2n)]}
\end{split}%
,
\end{equation}
and the resulting CE is presented in Fig. \ref{model4}.  

For the four models analysed we have started the CE numerical evaluation
considering $n=0.1$ with $0.1$ steps. The Models 1, 2 and 3 have presented
smaller CE values at the minimum $n$ adopted. In this cases, the models have
presented close CE minimum values. Model 4 presents the minimum CE at $%
n=0.8203$.\textcolor{black}{Unlike the other models, in this last case there is not a clear maximum value.}

\section{Tachyonic solutions}




For a given warp factor one can analyze if the scalar field $\phi$, which is
a solution of the Einstein equations, is real. Indeed, the relation that
makes such a discussion clear is given by Eq. (\ref{eqaphi}). We checked for
each model if the CE maximum and minimum points are related to tachyonic
solutions and the outcome is discussed in the following.

\textcolor{black}{If the CE global minimum coincides with the $n$ values that engender real
solutions, we have the CE method selecting real brane models. This is the
case for all the models analyzed in this work. We also analyzed the maximum
CE. We find that the maximum CE point for the Model 2 and 3 are
associated with a tachyonic solution. This could mimic a scenario with
unstable solutions decaying for a real stable one, located at the global CE
minimum.}

\textcolor{black}{In addition, by analyzing the energy-momentum tensor, we can note that for the models $1$, $2$ and $3$ the minimum CE selects positive definite energies, which are related to brane stable solutions. For the model $4$, $T^{00}$ presents regions with negative values, which is a drawback of model $4$ taking into account the physical description of  cosmological scenarios. For all models, the increasing of the CE yields the appearing of new maximum points, which can be related to the brane splitting. Once again, we note that the minimum CE is connected with the more favourable configurations.}

\section{Conclusions}

We have worked with the CE as a tool to select among many braneworld
scenarios described by a free parameter $n$. Considering the graphs
displaying the CE as a function of $n$, one can always identify a global
minimum CE. Moreover, some models present other higher entropy minimum
points, i.e. regions where $\partial (CE)/\partial n=0$ and $(\partial^2
CE)/(\partial n)^2 >0$, but that does not support the smaller CE minimum
value.

Another analysis was developed in order to verify the region where the brane
scalar field is a real function or a tachyonic one. This was performed by
the analysis of the sign of the warp factor's second derivative; if it is
positive, one has tachyonic solutions, otherwise it is real. For the $n$
values engendering real functions that are also the intervals for low
entropy global minimum, the CE is selecting real brane solutions.

We conclude that CE altogether with the warp factor's analysis could be a
physical observable to constraint one single free parameter $n$. On the
other hand, it does not exclude the possibility of indicating a more
intricate scenario where the scalar field could initially be at a potential
with higher CE value and then decay to smaller CE minima through symmetry
breaking mechanism or even a quantum tunneling process.

{It is worth to mention that the expected values for $n$ according to CE can
be further analysed in different sort of applications. For instance, in a
cosmological perspective, the Friedmann-like equations can be constructed
for the above braneworld scenarios. It can be checked if the values that CE
has indicated for $n$ in the different models are able to predict healthy
cosmological models, say, by describing a late-time accelerated expansion
universe \cite{perlmutter/1999}. }

On a more fundamental level, the predicted values for $n$ presented in this
work can be applied to attempt to recover the newtonian potential on a weak
gravitational field limit in the brane as in \cite{RSII}. Also, the possibility of
solving the hierarchy problem within these braneworld scenarios with the
referred values of $n$ can be further analysed (check \cite{randall/1999}). These applications have not been done so far once the braneworld models of Ref.\cite{branemariana} are quite recent. Moreover, with no restrictions for 
$n$, those applications could look artificial. In this way, in possession of
the here presented CE results, it is our intent to report on these subjects
soon.

\textbf{Acknowledgments}

MC and WTC thank Coordena\c{c}\~{a}o de Aperfei\c{c}oamento de Pessoal de N%
\'{\i}vel Superior (CAPES) and Conselho Nacional de Desenvolvimento Cient%
\'{\i}fico e Tecnol\'{o} gico (CNPq). RACC would like to thank S\~ao Paulo
Research Foundation (FAPESP), grant 2016/03276-5, for financial support.
PHRSM thank S\~ao Paulo Research Foundation (FAPESP), grant 2015/08476-0,
for financial support.

\appendix

\section{Deformation procedure}

Henceforth we fallow \cite{branemariana, alexroldao} in order to present a review of the deformation procedure calculation that can generate topological and non-topological defects described by single real scalar fields $\psi$. The mainly requirement in order to generate such solutions (known as BPS-like solutions \cite{BookB}) is to impose some features on the potentials $V$ that drives the scalar fields $\psi$. One demands that $V(\psi)$ generates a set of critical points, $\verb"{" \bar{\psi}_1,\ldots,\bar{\psi}_n\verb"}"$, such that $(dV/d\psi\vert_{\psi=\bar{\psi}_i}=0)$ as well as $V(\bar{\psi}_i)=0$, with $i=1,2,\ldots,n$. As a consequence, deformed potentials will engender two types of defects with different topologies: kinks (topological defects) if the potential has at least two degenerate minima, $\psi =\pm\bar{\psi}$; or lumps (non-topological defects) if the potential has one single minimum, $\psi=\bar{\psi}$.

A cyclic deformation method was developed in \cite{alexroldao} attempting to derive topological and non-topological solutions without the need to solve second order differential equations. A triggering primitive defect known \emph{a priori} departs the chain. Given that, one is able to construct an $N\hspace{-.1 cm}+\hspace{-.1 cm}2$-cyclic deformation chain (CDC) through the application of a chain rule constrained by hyperbolic or trigonometric fundamental relations. The initial scalar field solution was identified by the $\lambda \chi^4$ among other solutions in \cite{alexroldao}. 

Henceforth, we will describe the deformation procedure using the Greek letter $\chi$ for the triggering known solution and $\psi$ for the deformed defects. Note that, although these chains can generate kinks and lumps, have chosen \cite{alexroldao,alexmariana} only some of the lump solutions in order to build up the brane scenarios analyzed in this letter (as showed in Table II).

We will describe the $N+2$ chain evolved with the derivative of a generalized $\lambda$-deformation operation:
\begin{equation}
g^{[\lambda]}(\psi^{[\lambda]}) = \frac{d\psi^{[\lambda]}}{d\psi^{[\lambda-1]}},~~~\lambda =0,\, 1,\,2,\, \ldots,\,N.
\end{equation}
where the $\psi^{[\lambda]}$ are real scalar fields deformed from an $N\hspace{-.1 cm}+\hspace{-.1 cm}2$-CDC triggered by $\chi\equiv\chi(y)\sim \psi^{[-1]}$.

In \cite{alexroldao} they suggest the insertion of hyperbolic or trigonometric deformation functions in order to run the chain. Since the procedure is completely analogous for both kind of functions, we will report only the hyperbolic chain given by:
\begin{eqnarray}
\psi^{[0]}_{\chi} &=& \tanh{(\chi)}, \nonumber\\
\psi^{[1]}_{\chi} &=& \tanh{(\chi)}\sech{(\chi)}, \nonumber\\
\psi^{[2]}_{\chi} &=& \tanh{(\chi)}\sech{(\chi)}^{2}, \nonumber\\
&\vdots& \nonumber\\
\psi^{[N-1]}_{\chi} &=& \tanh{(\chi)}\sech{(\chi)}^{N-1}, \nonumber\\
\psi^{[N]}_{\chi} &=& \sech{(\chi)}^{N},
\label{AA}
\end{eqnarray}
where the subindex stands for the corresponding derivatives. Integrating these equations one has:
\begin{eqnarray}
\psi^{[0]}(\chi) &=& \ln{[\cosh{(\chi)}]}, \nonumber\\
\psi^{[1]}(\chi) &=& - \sech{(\chi)}, \nonumber\\
\psi^{[2]}(\chi) &=& - \frac{1}{2} \sech{(\chi)}^{2}, \nonumber\\
&\vdots& \nonumber\\
\psi^{[N-1]}(\chi) &=& - \frac{1}{N-1}\sech{(\chi)}^{N-1}, \nonumber\\
\psi^{[N]}(\chi) &=&  \sinh{(\chi)}\, _2F_1\left[\frac{1}{2},\, \frac{1+N}{2},\, \frac{3}{2},\, -\sinh{(\chi)}^2\right], \nonumber\\
\label{BB}
\end{eqnarray}
where the constants have been suppressed and $_2F_1$ is the Gauss' hypergeometric function.
From Eqs.~(\ref{AA}-\ref{BB}) it is possible to identifie that
\begin{equation}
g^{[\lambda]}(\psi^{[\lambda]}) = \sech{(\chi)} \equiv \exp{[-\psi^{[0]}]},
\end{equation}
with $\lambda = 1,\,2,\, \ldots,\,N-1$, so that $g^{[N]}(\psi^{[N]}) = 1/\sinh{(\chi)}$ and
\begin{equation}
\prod_{\lambda=1}^{N-1}{g^{[\lambda]}(\psi^{[\lambda]})} = \frac{d\psi^{[N-1]}}{d\psi^{[0]}} = \sech{(\chi)}^{N-1} \equiv \exp{[- (N-1) \psi^{[0]}]}.
\label{DD}
\end{equation}
From Eq.~(\ref{DD}), a complete expression for the chain rule of the $N\hspace{-.1 cm}+\hspace{-.1 cm}2$-CDC can be written as
\begin{equation}
\frac{d\psi^{[0]}}{d\chi}\,\prod_{\lambda=1}^{N}{g^{[\lambda]}(\psi^{[\lambda]})}\,\frac{d\chi}{d\psi^{[N]}} = 1,
\label{EE}
\end{equation}
from which $N\hspace{-.1 cm}+\hspace{-.1 cm}2$ deformation functions closing the cycles can be identified.

Considering generalized BPS functions \cite{BPS}, the potentials can be written as:
\begin{eqnarray}
y^{[\lambda]}_{\psi^{[\lambda]}} &=& \frac{d \psi^{[\lambda]}}{ds} = y^{[\lambda-1]}_{\psi^{[\lambda-1]}}\, g^{[\lambda]}(\psi^{[\lambda]}) = \nonumber \\
&=&y^{[\lambda-r]}_{\psi^{\lambda-r]}}\, g^{[\lambda-r+1]}(\psi^{[\lambda-r+1]})\nonumber\\
&=& w_{\chi} \psi^{[\lambda]}_{\chi},
\end{eqnarray}
with $w_{\chi} = d\chi/ds$.
In this case, a constraint equation can be found:
\begin{eqnarray}
&&\sum_{\lambda=0}^{N-1}{(y^{[\lambda]}_{\psi^{[\lambda]}})^2}=(y^{[0]}_{\psi^{[0]}})^2 \sum_{\lambda=0}^{N-1}{\left(\frac{d\psi^{[\lambda]}}{d\psi^{[0]}}\right)^2}= \nonumber\\
&=&  w_{\chi}^2 \sum_{\lambda=0}^{N-1}{(\psi^{[\lambda]}_{\chi})^2}
= w_{\chi}^2 \tanh{(\chi)}^2\,\sum_{\lambda=0}^{N-1}{(\sech{(\chi)})^{2\lambda}}=\nonumber\\
&=&  w_{\chi}^2 \tanh{(\chi)}^2\,\frac{1 - \sech{(\chi)}^{2N}}{1-\sech{(\chi)}^2}=\nonumber\\
&=& w_{\chi}^2 \left[ 1 - (\psi^{[N]}_{\chi})^2 \right] = w_{\chi}^2 - (y^{[N]}_{\psi^{[N]}})^2,
\label{BBBB}
\end{eqnarray}
which isolating $w_{\chi}^2$ gives:
\begin{eqnarray}
\sum_{\lambda=0}^{N}{(y^{[\lambda]}_{\psi^{[\lambda]}})^2} &=& w_{\chi}^2.
\label{cd1}
\end{eqnarray}

From Eq~(\ref{cd1}) Bernardini and Rold\~ao were able to write the following mass relation:
\begin{eqnarray}
\sum_{\lambda=0}^{N}{(M^{\psi^{[\lambda]}})} &=& M^{\chi},
\label{cd2}
\end{eqnarray}
where the mass $M$ is defined as the integration of the localized energy densities $\rho(\psi)=y_{\psi}^2$ and $\rho(\chi)=w_{\chi}^2$ over the extra coordinate, $y$. This energy density localization pattern is an essencial requirement for the configurational entropy calculation effectuated in the present work.

Giving this recipe for CDC, \cite{alexroldao} show that it is also possible to insert trigonometric deformation functions by changing $\tanh{(\chi)}$ by $-\sin{(\chi)}$ and $\sech{(\chi)}$ by $\cos{(\chi)}$ into Eqs.~(\ref{AA}-\ref{BB}). As a consequence, it will generate a set of constraints analogous to those described by Eqs.~(\ref{cd1}-\ref{cd2}).

Having established the CDC procedure, we presente the two primitive kink solutions, among those used to trigger the $N+2$-CDC in \cite{alexroldao}, which generate results that are from direct interest in this letter:
\begin{eqnarray}
\label{chil}
\chi_a(y)&=&\tanh(y), \\
\chi_b(y)&=&\sech(y).
\end{eqnarray}
Deformed defects, here named $\psi_1, \psi_2, \psi_3$ and $\psi_4$, derived in \cite{alexroldao} from $\chi_a$ and $\chi_b$ hold a lump-like behavior. In \cite{branemariana} they were identified with $A(y)$ from Eq.~(\ref{A2}), as to support warp factors engendering consistent braneworld scenarios guided by a real scalar field $\phi$ .

In Table II we present the expressions for these four chosen lumps. The CDC was triggered by the known solution $\chi_a$ in order to derive $\psi_1$, $\psi_2$ and $\psi_3$. The deformed lump $\psi_4$, was derived considering the triggering function $\chi_b$.

\begin{table}[h!]
\centering
\begin{tabular}{| L{2cm} | L{4cm} | L{5cm} |}
\hline
{Chain} & Deformation function &Deformed lump solution\\
\hline
{3-cyclic} & Hyperbolic &$\psi_1=\frac{-\ln[\cosh[n \tanh(y)] \sech(n)]}{n}$\\
{3-cyclic} &Trigonometric&$\psi_2=\frac{\cos[n \tanh(y)] - \cos(n)}{n}$\\
\hline
{4-cyclic} &Hyperbolic&$\psi_3 =\frac{(\sech[n \tanh(y)] - \sech(n))}{n}$\\
{4-cyclic}&Trigonometric&$\psi_4 = \frac{(2 n \sech(y) + \sin[2 n \sech(y)])}{4 n}$\\
\bottomrule[1.5pt]
\end{tabular}
\caption{Lump-like deformed defects derived with hyperbolic and trigonometric functions in a CDC from \cite{alexroldao}.}
\label{tabela}
\end{table}

\end{document}